\RequirePackage[hyphens]{url}
\documentclass[journal=jpccck,manuscript=article]{achemso}

\usepackage[version=3]{mhchem} 
\usepackage{xcolor}
\usepackage{soul} 
\usepackage{amssymb}
\usepackage{lineno,hyperref}
\usepackage{graphicx}
\usepackage{dcolumn}
\usepackage{bm}
\usepackage{float}
\usepackage{lipsum}

\newcommand{\DIPC}[0]{
Donostia International Physics Center (DIPC),
Paseo Manuel de Lardizabal 4, 20018 Donostia-San Sebasti\'an, Spain}
\newcommand{\CFM}[0]{
Centro de F\'{\i}sica de Materiales CFM/MPC (CSIC-UPV/EHU), Paseo Manuel de Lardizabal 5, 20018 Donostia-San Sebasti\'an, Spain}
\author{Auguste Tetenoire}
\email{auguste.tetenoire@dipc.org}
 \affiliation{\DIPC}

\author{J.\ I\~naki Juaristi}
\email{josebainaki.juaristi@ehu.eus}
\affiliation{Departamento de Pol\'{\i}meros y Materiales Avanzados: F\'{\i}sica, Qu\'{\i}mica y Tecnolog\'{\i}a, Facultad de Qu\'{\i}micas (UPV/EHU), Apartado 1072, 20080 Donostia-San Sebasti\'an, Spain}
\alsoaffiliation{\CFM}
\alsoaffiliation{\DIPC}

\author{Maite Alducin}
\email{maite.alducin@ehu.eus}
 \affiliation{\CFM}
 \alsoaffiliation{\DIPC}

\title{Photo-Induced CO Desorption Dominates over Oxidation on Different O+CO Covered Ru(0001) Surfaces}

\date{\today}

\begin{document}

\begin{tocentry}

\includegraphics[width=1\columnwidth]{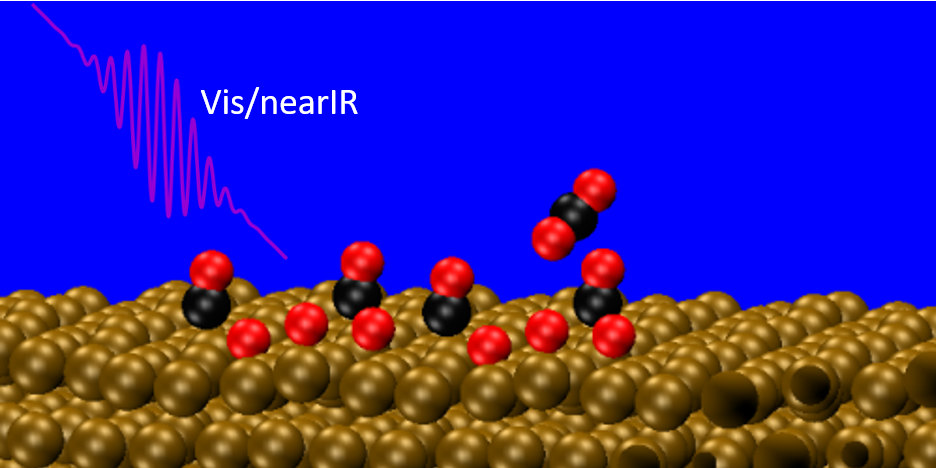}

\end{tocentry}

\begin{abstract}\label{sec:abstract}
The photo-induced desorption and oxidation of CO on Ru(0001) is simulated using ab initio molecular dynamics with electronic friction that accounts for the non-equilibrated excited electrons and phonons. Different (O,CO) coverages are considered, the experimental room temperature coverage consisting in 0.5ML-O+0.25ML-CO (low coverage), the saturation coverage achieved experimentally at low temperatures (0.5ML-O+0.375ML-CO, intermediate coverage), and the equally mixed monolayer that is stable according to our calculations but not experimentally observed yet (0.5ML-O+0.5ML-CO, high coverage). The results of our simulations for the three coverages are consistent with femtosecond laser experiments showing that the CO photo-desorption largely dominates over CO photo-oxidation. These results cannot be explained in terms of the distinct activation energies calculated for the relaxed surfaces. Different (dynamical) factors such as the coupling to the laser-excited electrons and, more importantly, the interadsorbate energy exchange and the strong surface distortions induced in the more crowded surfaces are fundamental to understand the competition between these two processes under the extremely non-equilibrated conditions created by the laser. 
\end{abstract}



\section{Introduction \label{sec:intro}}
Understanding the oxidation of CO on transition metal surfaces is one of the long-standing important problems in surface science that continues motivating extensive fundamental research~\cite{spronsen17,Henss2019Feb,zhou19,cueto20,jin22,larue22,Tiwari22}. A particularly interesting surface is Ru(0001) with coadsorbed CO and atomic O. It has been demonstrated that under ultra high vacuum (UHV) conditions the oxidation process cannot be thermally activated~\cite{Kostov1992Nov,bonn99}. Interestingly, the reaction can be propelled by exciting the system with electromagnetic radiation~\cite{bonn99,obergJCP2015,ostrom2015Feb,Larue2015Jul}. Specifically, this has been achieved by using near-infrared femtosecond laser pulses that excite the electrons of the system. Subsequently, those laser-excited electrons transfer energy directly to the adsorbates, but also indirectly, via excited phonons that result from electron-phonon coupling. Under these conditions, both CO and CO$_2$ are desorbed from the surface. 

A proper characterization of this kind of experiments requires an accurate simulation of the reaction dynamics. In a successful approach to this problem, the response of the metal to the laser pulse is described in terms of coupled electronic and phononic excitations characterized, respectively, by time-dependent electronic ($T_e$) and phononic ($T_l$) temperatures~\cite{anisimov74}. Subsequently, simulations of the adsorbates dynamics in this highly excited system are performed, accurately describing the adsorbate-surface interaction with potential energy surfaces based on density functional theory (DFT)~\cite{vazha09,fuchsel10,fuchsel11,loncaricprb16,scholz16}. In this respect, the development of the ab initio molecular dynamics with electronic friction method (AIMDEF)~\cite{novkoprb15,novkoprb16,novkoprb17}, and in particular its version that incorporates electronic and phononic temperatures (($T_\textrm{e}, T_\textrm{l}$)-AIMDEF), has allowed to treat the multidimensional nonadiabatic dynamics of this kind of photo-induced reactions at the DFT level taking into account the coupling of the adsorbates to only the excited electrons in some cases~\cite{juaristiprb17} and to both the excited electronic and phononic systems in others~\cite{alducin2019,scholz19,tetenoire2022}. Also the coadsorbate energy exchange that may become important as surface coverage increases~\cite{denzler03,xin15} can be readily included in this kind of simulations~\cite{juaristiprb17,alducin2019}. The ($T_\textrm{e}, T_\textrm{l}$)-AIMDEF scheme is of special relevance for CO oxidation at photo-excited metal surfaces due to the high dimensionality involved in the reaction. In addition to the minimal 9 dimensions of the reacting O and CO, the mentioned coadsorbate energy exchange, energy exchange of the adsorbates (particularly CO) with the surface atoms, and surface distortions caused not only by the excited phonons, but also by the excited adsorbates, including possible formation of subsurface O absorbates at high surface temperatures~\cite{bottcher00,cai16}, are important factors that must be included in a realistic simulation.

The preparation of the adsorbate layer in the experiments of refs.~\citenum{bonn99,obergJCP2015,ostrom2015Feb} is performed in the following way. First, the Ru(0001) surface is dosed with oxygen up to saturation which consists in a 0.5~ML O coverage~\cite{Kostov1992Nov}. In a second step, the dosing of the surface with CO is produced also up to saturation. Theoretical studies motivated by these experiments, devoted to the description of the energetics of the reaction path~\cite{ostrom2015Feb} and to the use of kinetic models~\cite{obergJCP2015} to understand the measured desorption yields, assumed a 0.25 ML CO coverage. Also, in a recent work, the ($T_\textrm{e}, T_\textrm{l}$)-AIMDEF method has been used for the first time to simulate the femtosecond laser pulse induced oxidation of CO in the Ru(0001) surface for this coverage~\cite{tetenoire2022}. In that work, the simulations reproduced and allowed to understand the large experimental branching ratio between desorption and oxidation that exceeds one order of magnitude, in spite of the fact the CO oxidation is energetically favored over CO desorption at that 0.5~ML O + 0.25~ML CO surface coverage. Nevertheless, though this is the reported saturation coverage at room temperature~\cite{Kostov1992Nov}, in the experiments the temperature was lowered before the laser excitation to 100~K~\cite{bonn99,obergJCP2015,ostrom2015Feb}, for which a higher CO saturation coverage cannot be excluded~\cite{Kostov1992Nov}.

The above discussed facts constitute our motivation to perform ($T_\textrm{e}, T_\textrm{l}$)-AIMDEF for different surface coverages. More precisely, we have simulated the photo-induced CO desorption and oxidation on three different O+CO covered Ru(0001) surfaces, for which the equilibrium structures and minimum energy reaction paths for CO desorption and oxidation were studied and characterized using DFT in a previous work~\cite{tetenoire2021}. In the three cases, the O coverage is fixed to 0.5~ML and only the CO coverage, which is not clearly determined in the femtosecond laser experiments, is different. We will denote these CO coverages as low, intermediate, and high coverages following the nomenclature of ref.~\citenum{tetenoire2021}. The low coverage consisting in the 0.5~ML O + 0.25~ML CO constitutes the CO saturation coverage at 300~K~\cite{Kostov1992Nov}. The intermediate coverage is formed by 0.5~ML O + 0.375~ML CO and it is close to the CO saturation coverage reported at temperatures below 120~K~\cite{Kostov1992Nov}. Finally, the high coverage defined by 0.5~ML O + 0.5~ML CO, although not experimentally observed, is energetically stable according to our previous DFT calculations~\cite{tetenoire2021} and hence useful to understand the coverage dependence of the photo-induced desorption and oxidation of CO. 

The paper is organized as follows. The theoretical methods and computational settings used to model the different (O,CO)-covered Ru(0001) surfaces are described in the Methods section. The results of the photo-induced oxidation and desorption dynamics, obtained from our ($T_\textrm{e}, T_\textrm{l}$)-AIMDEF simulations for each of the three coverages considered, are presented and comparatively discussed in the Results and Discussion section. The main conclusions extracted from our dynamics simulations are summarized in the Conclusions section.

\section{Theoretical Methods \label{sec:methods}}

\subsection{Photo-Induced Desorption Model\label{sec:theory}}

The photo-induced desorption and oxidation of CO from the (O,CO)-covered Ru(0001) surface is simulated with ab initio classical molecular dynamics using the ($T_\textrm{e}, T_\textrm{l}$)-AIMDEF methodology~\cite{alducin2019}. These kinds of simulations incorporate the effect of the laser-induced excited electrons in both the adsorbates dynamics and the surface atoms dynamics. An extensive description of this method and its applicability to model femtosecond laser-induced reactions on metals can be found in refs.~\citenum{alducin2019,tetenoire2022}. Hence, only its main ingredients will be reviewed next. 

As a first step, the response of the metal surface to the near-infrared laser pulse is described within the two-temperature model (2TM)~\cite{anisimov74}. In this model, the laser-induced electronic and ensuing (electron-induced) phononic excitations are represented by two coupled heat thermal baths, being  $T_\textrm{e}(t)$ and $T_\textrm{l}(t)$ their associated time-dependent temperatures obtained from the following equations:
\begin{eqnarray}\label{eq:2tm}
C_{e}\frac{\partial T_{e}}{\partial t}& =& \frac{\partial}{\partial
z}\kappa\frac{\partial T_{e}}{\partial z}-g\,(T_{e}-T_{l})+S(z,t) \, ,
\nonumber \\
C_{l}\frac{\partial T_{l}}{\partial t} & =& g(T_{e}-T_{l}) \, ,
\end{eqnarray}
where $C_{e}$ is the the electron heat capacity, $C_{l}$ is the phonon heat capacity, $\kappa$ is the electron thermal conductivity, $g$ is the electron-phonon coupling constant, $z$ the perpendicular position relative to the surface and $S(z,t)$ is the absorbed laser power per unit volume that depends on the shape, wavelength, and fluence of the applied pulse. 

Once $T_{e}(t)$ and $T_{l}(t)$ are known, the effect of the laser-excited electrons on each adsorbate is described through the following Langevin equation:
\begin{equation}\label{eq:langevin}
m_i\frac{d^2\mathbf{r}_i}{dt^2}=-\nabla_{\mathbf{r}_i} 
V(\mathbf{r}_1,...,\mathbf{r}_N)-\eta_{e,i}(\mathbf{r}_i)\frac{d\mathbf{r}_i}{dt} 
+\mathbf{R}_{e,i}[T_{e}(t),\eta_{e,i}(\mathbf{r}_i)] \, , 
\end{equation}
where $m_i$, $\textbf{r}_i$, and $\eta_{e,i}$ are the mass, position vector, and electronic friction coefficient of the $i^{th}$ atom conforming the set of adsorbates. The first term in the right hand side of the equation is the adiabatic force that depends on the position of all (adsorbates and surface) atoms. The effect of the electronic excitations and deexcitations on the adsorbates is described by the electronic friction force (second term) and the electronic stochastic force (third term), both related through the fluctuation-dissipation theorem. In particular, $\mathbf{R}_{e,i}$ is modeled by a Gaussian white noise with variance
$\mathrm{Var}[\mathbf{R}_{e,i}(T_{e},\eta_{e,i})]=(2 k_B T_{e}(t)
\eta_{e,i}(\mathbf{r}_i))/\Delta t $,
where $k_B$ and $\Delta t$ are the Boltzmann constant and the time-integration step, respectively. For each atom $i$, the electronic friction coefficient $\eta_{e,i}(\mathbf{r}_i)$ is calculated with the local density friction approximation (LDFA)~\cite{juaristi08,alducinpss17} using the Hirshfeld partitioning scheme~\cite{Hirshfeld1977} to extract on-the-fly the bare surface electron density (see refs.~\citenum{novkoprb15,novkoprb16} for details).

Furthermore, the heating of the surface lattice due to the laser-induced electronic excitations, which is described through $T_\textrm{l}(t)$, is assured by coupling the surface atoms in the two upper layers to the Nos\'{e}-Hoover thermostat~\cite{Nose84,Hoover1985}. Hence, these surface atoms with mass $m_j$ and position vector $\mathbf{r}_j$ are subjected to the following equations of motion,
\begin{align}
m_j \frac{d^2\mathbf{r}_j}{dt^2}&=-\nabla_{j}V({\bf r}_1,...,{\bf r}_N)
-m_j \, \xi \, \frac{d\mathbf{r}_j}{dt} \, , \label{eq:nh1}\\
\frac{d\xi}{dt}&=\frac{1}{Q}\left(\sum_{j} m_j
\left| \frac{d\mathbf{r}_j}{dt}\right|^2-3 N k_B
T_{l} \right),\label{eq:nh2}
\end{align}
where $N$ is the number of atoms coupled to the thermostat (in our case the first two layers), $Q$ is a parameter with dimensions of energy$\times$time$^2$ that acts as the mass of the dynamical variable $s$ with associated conjugated momentum $p_s$, and $\xi=Q^{-1} s p_s$ is the thermodynamic friction coefficient~\cite{Huenenberger2005}. Finally, the fourth and fifth Ru layers are kept frozen, while the third layer is used as a transition layer to simulate that the lattice heating flows from the surface to the bulk. Hence, the movement of the third layer atoms is described by the classical Newton equations of motion and the adiabatic approximation.
The ($T_\textrm{e}, T_\textrm{l}$)-AIMDEF method and similar methods that combine the 2TM and the Langevin dynamics for the adsorbates have been widely used to simulate the femtosecond laser induced dynamics and reactions of adsorbates at metal surfaces~\cite{tullyss94,springer96,vazha09,fuchsel10,fuchsel11,loncaricprb16, loncaricnimb16, scholz16, juaristiprb17,alducin2019, scholz19}.  

\subsection{ General DFT Computational Settings}
\begin{figure}   \includegraphics[width=0.75\columnwidth]{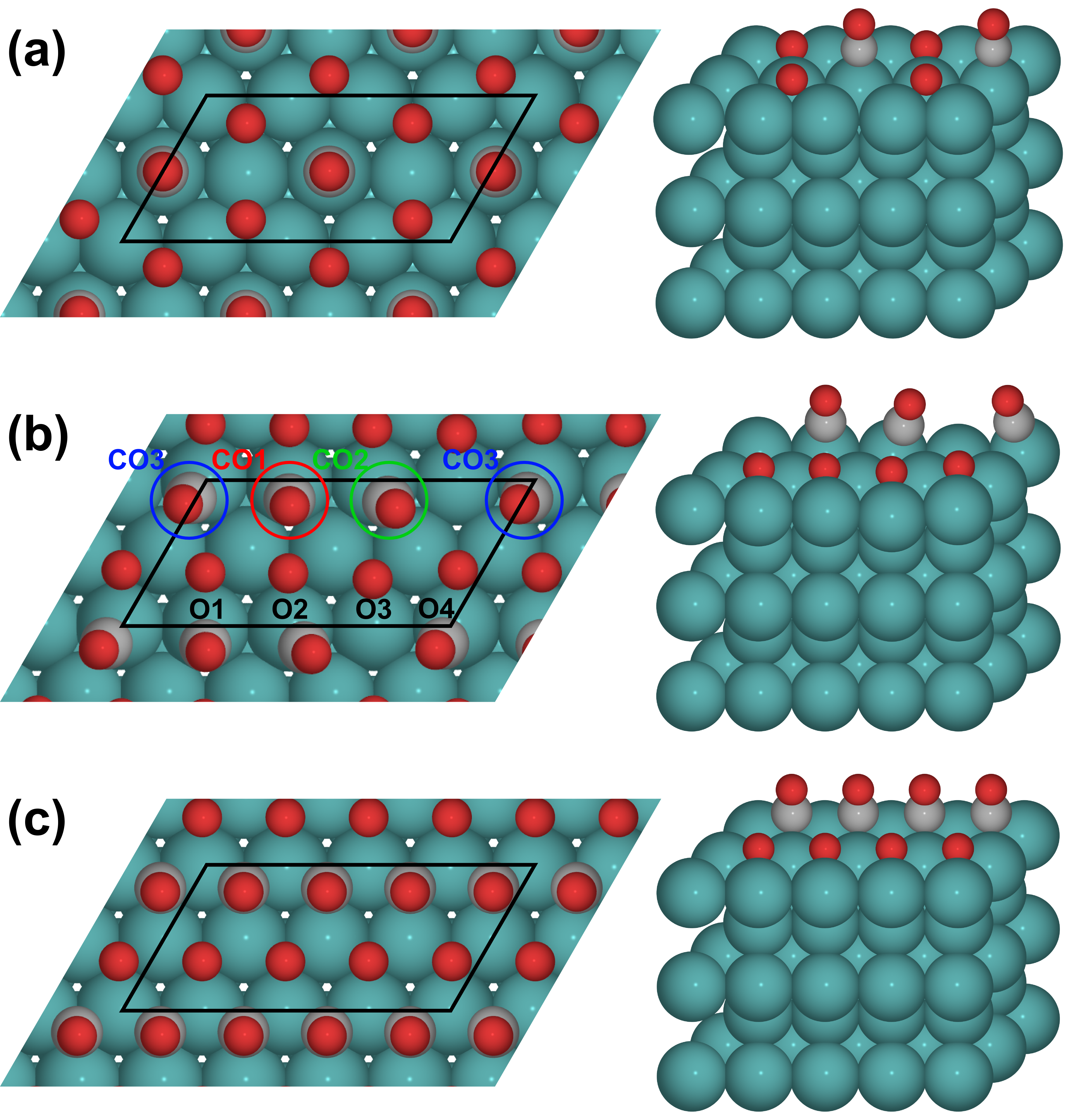}
  \caption{Top (left panels) and perspective (right panels) views of the energetically favored structures found in ref.~\citenum{tetenoire2021} for Ru(0001) covered by: (a) 0.5~ML O + 0.25~ML CO, (b) 0.5~ML O + 0.375~ML CO, and (c) 0.5~ML O + 0.5~ML CO. The black parallelograms depict the (4$\times$2) surface unit cell used in the AIMDEF calculations for each coverage. Color code: O atoms in red, C in gray, and Ru in blue. For clarity, the periodic images of the O and CO adsorbates in the perspective view are not shown. Images prepared with ASE~\cite{ase}.
  }
   \label{fig:cell}
\end{figure}

All ($T_\textrm{e}, T_\textrm{l}$)-AIMDEF simulations are performed with {\sc vasp}~\cite{vasp1,vasp2} (version 5.4) and the AIMDEF module~\cite{blanco14,saalfrank14,novkoprb15,novkoprb16,novkonimb16,novkoprb17,juaristiprb17} using for each coverage the corresponding optimized supercell shown in Figure~\ref{fig:cell}. Each periodic supercell is defined by a (4$\times$2) surface unit cell and a vector length along the surface normal of 30.22~{\AA}. Within this supercell, the covered Ru(0001) surface is described by five layers of Ru atoms and the specific mixed (O,CO) overlayer. The Ru topmost layer is separated from the bottom of the periodic Ru slab by about $19$~{\AA} of vacuum.  The employed (4$\times$2) surface cell containing various adsorbates permits a reliable description of the interadsorbates interactions, which are expected to affect the adsorbates dynamics at sufficiently large coverages~\cite{denzler03,xin15,sung16,juaristiprb17,alducin2019, serrano2021}. As shown in the figure, in the low coverage (0.5ML~O+0.25ML~CO), each CO adsorbs atop a Ru atom, while the O atoms occupy the second nearest hcp and fcc sites forming a honeycomb arrangement around CO. In the optimized intermediate coverage, the O atoms adsorb at hcp sites in a p(1$\times$2) arrangement and the CO molecules occupy the empty space left between the O arrays. Specifically, the CO molecules labeled as CO2 and CO3 in the figure are located in the line joining the top and bridge sites, with their axes slightly tilted towards the bridge site. The molecule labeled CO1 is located in the line joining the top and fcc sites and is slightly tilted towards the fcc site. In the O+CO saturated Ru(0001) surface (high coverage), both the O and CO adsorb on hcp sites forming two inserted p(1$\times$2) structures. 

During the ($T_\textrm{e}, T_\textrm{l}$)-AIMDEF simulations, the adiabatic forces are calculated with non spin-polarized DFT using the van der Waals exchange-correlation functional proposed by Dion \textit{et al.}~\cite{Dion2004} and the same computational parameters that were used in our previous study on the energetics of CO desorption and oxidation at different coverages~\cite{tetenoire2021}. Specifically, the electronic ground state is determined at each integration step by minimizing the system total energy up to a precision of $10^{-6}$~eV. In doing so, integration in the Brillouin zone is performed using a $\Gamma$-centered 3$\times$6$\times$1  Monkhorst-Pack grid of special \textbf{k} points~\cite{Monkhorst1976} and the Methfessel and Paxton scheme of first order with a broadening of 0.1~eV to describe partial occupancies of each state~\cite{Methfessel1989}. The latter are expanded in a plane-wave basis set with an energy cut-off of $400$~eV, whereas the electron-core interaction is treated with the projected augmented wave (PAW) method~\cite{Blochl1994Dec} that is implemented in VASP~\cite{Kresse1999}. All dynamics simulations are performed with the Beeman integrator implemented in our AIMDEF module~\cite{blanco14} using a time step of 1~fs. Each trajectory is propagated up to 4~ps. As possible outcomes, we observe CO desorption and CO oxidation (i.e., the O+CO recombinative desorption). Atomic O is strongly adsorbed and no desorption events are obtained. Oxygen subsurface migration is  however temporarily possible, being more probable the higher the coverage. 

\subsection{Initial Conditions}
For each coverage, a preliminary simulation is performed in which the three outmost Ru layers are equilibrated at 100~K during 20~ps using the Nos\'e-Hoover thermostat~\cite{Nose84,Hoover1985} [Eqs.~(\ref{eq:nh1}) and~(\ref{eq:nh2})], while the mixed (O,CO) adlayer follows an adiabatic dynamics. The adlayer interchanges energy with the thermalized Ru(0001) surface and becomes progressively thermalized at the end of this preliminary simulation. The initial positions and velocities of the moving atoms (i.e., the adlayer and the three topmost Ru layers) are randomly taken from the data generated in the last 5~ps of the thermalization trajectory. We have verified that the selected trajectories (between 150-200, depending on coverage) reasonably reproduce the expected Maxwell-Boltzmann distribution.

\subsection{Calculation of observables}
A molecule (CO or CO$_2$) is counted as desorbed, if its center of mass crosses the plane $z$=6.5~{\AA}, measured respect to the topmost Ru surface layer, with positive velocity along the surface normal. Since the Ru atoms are also moving, the reference $z$ plane is calculated at each instant $t$ as the mean height of the Ru atoms in the topmost layer. The CO desorption and oxidation probabilities per CO molecule are calculated for each coverage as
\begin{equation}
    P_\textrm{des}(A)=\frac{N_\textrm{des}(A)}{N_\textrm{t} N_\textrm{CO}}
\end{equation}
where $N_\textrm{des}(A)$ is the number of desorbing molecules (CO or CO$_2$), $N_\textrm{t}$ is the total number of trajectories, and $N_\textrm{CO}$ is the number of CO molecules in the simulation cell (2, 3, and 4 for low, intermediate, and high coverages, respectively). In the intermediate coverage, the site-resolved desorption probabilities refers to the desorption probabilities for each of the three distinct CO molecules in the cell, i.e., $P_\textrm{des}$(CO1), $P_\textrm{des}$(CO2), and $P_\textrm{des}$(CO3).

The mean total kinetic energy $\langle E_\textrm{kin}\rangle (t)$ and mean center-of-mass kinetic energy $\langle E_\textrm{cm}\rangle (t)$ per adsorbate type (i.e., O or CO) are calculated at each instant $t$ as an average over all the trajectories and all the atomic or molecular species under consideration.

\section{Results and Discussion\label{sec:results}}
All the simulations correspond to irradiating the surface with a 800~nm Gaussian pulse of 110~fs duration and absorbed fluence F=200~J/m$^2$ as in experiments~\cite{bonn99}. The corresponding $T_\textrm{e}(t)$ and $T_\textrm{l}(t)$ curves calculated with Eq.~\eqref{eq:2tm} (input parameters for the Ru slab as in refs.~\citenum{vazha09,scholz16,juaristiprb17,tetenoire2022}) are shown in Figure~\ref{fig:2tm}. The CO desorption and CO oxidation probabilities obtained from the ($T_\textrm{e}, T_\textrm{l}$)-AIMDEF simulations for each coverage are summarized in Table~\ref{tab:table1}.  Starting with the CO desorption process, the largest to the lowest probabilities correspond to the intermediate, high, and low coverages. As also shown in the table, this ordering agrees with the values of the CO desorption barriers obtained in our previous DFT analysis of the coverage-dependent energetics of the two reactions~\cite{tetenoire2021}. The photo-oxidation process also occurs on the three coverages. The highest probability for this reaction is unexpectedly obtained for the high coverage, i.e., the coverage for which the activation barrier for CO oxidation is the largest. The analysis of the oxidizing trajectories will allow us to rationalize this result. Compared to the CO desorption process, the CO oxidation probabilities are considerably lower in the three cases. In spite of having a limited statistics for CO oxidation, let us remark that the resulting large branching ratios between desorption and oxidation are in line with the experimental observations, P$_\textrm{des}$(CO)/P$_\textrm{des}$(CO$_2$)$=35$~\cite{bonn99} and 31~\cite{obergJCP2015}. As shown in ref.~\citenum{tetenoire2022} for the low coverage, the extremely small P$_\textrm{des}$(CO$_2$) values are due to the very reduced configurational space leading to CO oxidation as compared to that for CO desorption. More precisely, at low coverage, we found that oxidation is unlikely due to the difficult access to the transition state region that requires the O atom crossing the bridge site and finding the CO conveniently close and tilted to form the CO$_2$ molecule and due also to the fact that this access does not guarantee a successful recombination. In this work, we will hence focus in understanding how the CO desorption and CO oxidation dynamics depend on the Ru(0001) coverage.
\begin{table*}
\caption{\label{tab:table1} ($T_\textrm{e}, T_\textrm{l}$)-AIMDEF CO desorption probability $P_\textrm{des}$(CO), CO oxidation probability $P_\textrm{des}$(CO$_2$), and CO to CO$_2$ branching ratio calculated for an absorbed fluence F=200~J/m$^2$ and three different (O,CO)-Ru(0001) coverages. $N_\textrm{t}$ is the number of trajectories run for each coverage. Activation energies for CO desorption $E_\textrm{TS}$(CO) and CO oxidation $E_\textrm{TS}$(CO$_2$) are from ref~\citenum{tetenoire2021}.}
\begin{tabular}{ccccccc}
\hline 
Coverage & $N_\textrm{t}$&$P_\textrm{des}$(CO) & $P_\textrm{des}$(CO$_2$) & ratio &$E_\textrm{TS}$(CO) & $E_\textrm{TS}$(CO$_2$)\\ \hline
0.5ML~O+0.25ML~CO & 200 & 18.25\% &  0.5\% & 36.5 &1.57 & 1.19\\
0.5ML~O+0.375ML~CO & 199 & 45.06\% &  0.67\% & 67.3& 0.58,0.73& 0.80\\
0.5ML~O+0.5ML~CO & 139 & 34.53\% &  1.26\% & 27.4 & 0.88 & 2.01\\

\end{tabular}
\end{table*}

\subsection{CO Desorption Dynamics}
\begin{figure}
\includegraphics*[angle=0,width=0.5\columnwidth]{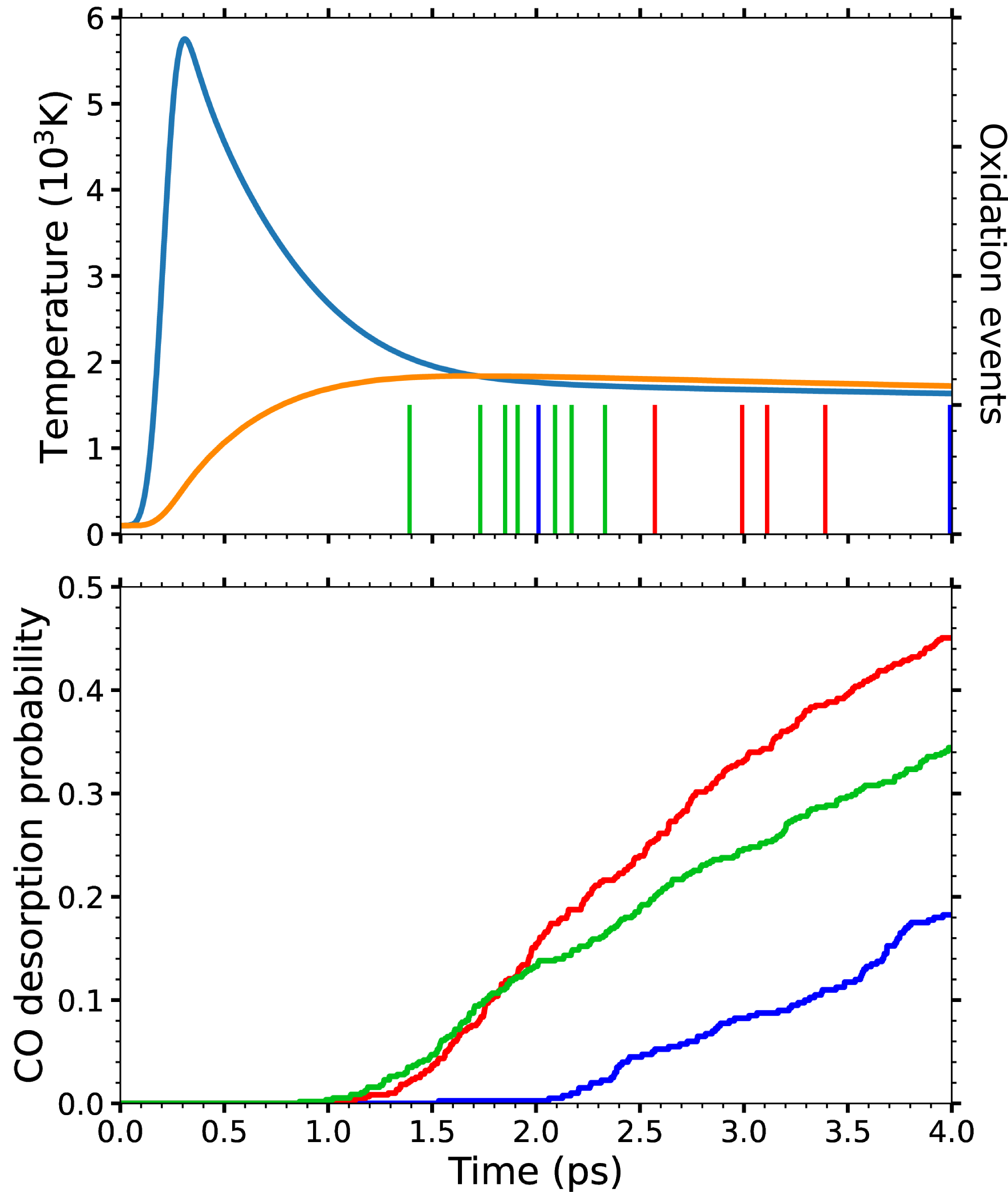}%
\caption{Top: 2TM-calculated electronic (blue) and lattice (orange) temperatures induced by a 800~nm laser pulse with 120~fs FWHM and $F$=200~J/m$^2$ (left $y$-axis). The peak of the laser pulse is at 236~fs. Histograms showing the instants at which the CO oxidation events occur for low (blue), intermediate (red), and high (green) coverages (right $y$-axis). The bin width is 20~fs. Bottom: CO photo-desorption probability against time for O+CO-Ru(0001) with the low (blue), intermediate (red), and high (green) coverages.}
\label{fig:2tm}
\end{figure}

The time evolution of the CO desorption probabilities $P_\textrm{des}(t)$ for the three coverages are compared in Figure~\ref{fig:2tm} (bottom panel). Desorption starts at about 1~ps in the intermediate and high coverages and in both cases the desorption probability increases monotonously with time from these very first instants. In contrast, the first desorption events in the low coverage occurs about 0.5~ps later and it is not until $t$=2~ps that the desorption probability starts to increase with time. A rough comparison among the slopes of the three $P_\textrm{des}(t)$ curves suggests a faster CO desorption rate in the intermediate coverage (red) as compared to the rather similar rates obtained in the low (blue) and high (green) coverages. The analysis of different properties of the adsorbates dynamics, such as kinetic energy, coupling to the hot electrons, and interadsorbate energy exchange, allows us to understand these results. 
%

\begin{figure}[t]
\includegraphics*[width=0.5\textwidth]{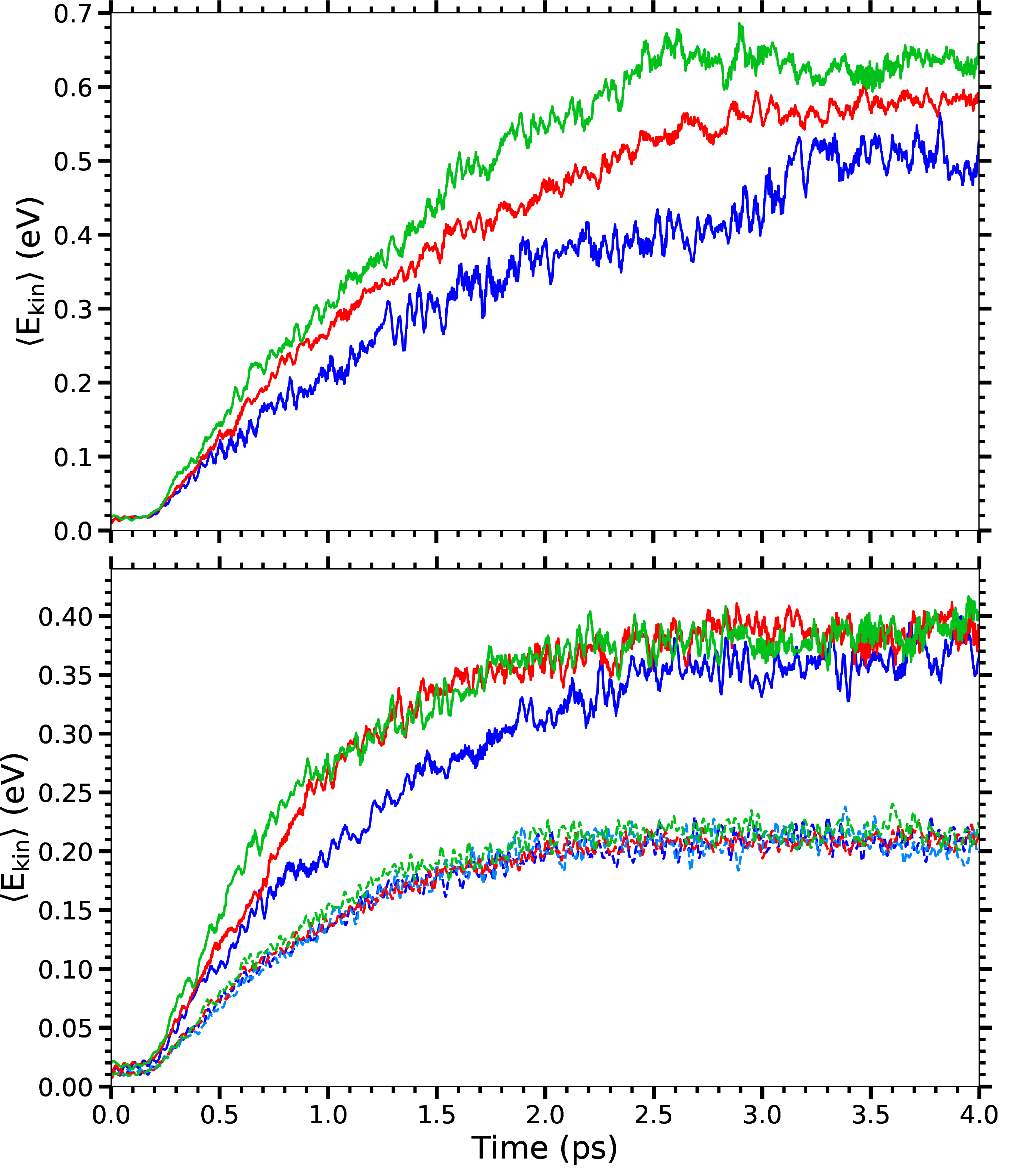}%
\caption{Coverage dependence of the time evolution of the mean total kinetic energy of desorbing CO (top panel), non-desorbing CO (bottom panel, solid lines) and adsorbed O (bottom panel, dashed lines). Results for low coverage in blue, intermediate coverage in red, and high coverage in green. 
}
\label{fig:ekin_coverage}
\end{figure}
The mean kinetic energy $\langle E_\textrm{ekin} \rangle$ of the desorbing and non-desorbing CO are represented in the top and bottom panels of Figure~\ref{fig:ekin_coverage}, respectively. In both cases, the energy uptake during the first 0.4~ps is similar in the low and intermediate coverages, but slightly larger for the high coverage. During this initial stage, the main mechanism contributing to the adsorbates excitation is the coupling to the hot electrons (compare $T_\textrm{e}$ to $T_\textrm{l}$ in Figure~\ref{fig:2tm}). Since CO adsorbs on hollow sites in the high coverage case (i.e., very close to the surface) and at near-top and top sites in the intermediate and low coverage cases, respectively, this coupling is stronger in the former than in the two latter, explaining the slightly larger initial energy gain in the high coverage. This larger energy gain may also explain that $P_\textrm{des}$(CO) is initially larger for this coverage and not for the intermediate one, which has a lower activation energy for desorption (see Table~\ref{tab:table1}). There are various interesting features when comparing the mean kinetic energy among the three coverages. In the case of the desorbing CO, $\langle E_\textrm{kin} \rangle$ is the largest during the whole integration interval for the high coverage. The same behavior is observed for the non-desorbing CO but only in the time interval $t<$1~ps. At longer times the non-desorbing CO molecules in the high and intermediate coverages have rather equal $\langle E_\textrm{kin} \rangle$ values, indicating that in both cases the adsorbates are already thermalized with the surface. In the low coverage, both the desorbing and non-desorbing CO exhibit a slower energy gain (and smaller $\langle E_\textrm{kin} \rangle$) as compared to the results obtained in the other coverages. There are two factors that can contribute to the dependence of the energy gain on coverage, the mentioned coupling to the hot electrons given by the friction coefficient $\eta_{e,i}$ in eq.~\eqref{eq:langevin} and the interadsorbate energy exchange, which is expected to be relevant at large CO coverages~\cite{xin15,alducin2019}. Since the CO molecules stay nearby their initial top positions during the first 1.0~ps in the low coverage, the coupling to the hot electrons is the weakest. Specifically, the CO mean friction coefficients $\langle \eta_e\rangle(t)$ are $\sim$0.09, 0.10, and 0.13-0.10~a.u. (atomic units) for the low, intermediate, and high coverages, respectively, within this time interval. However, such differences in the coupling, in particular between the low and intermediate coverages, seem insufficient to explain the slower energy uptake obtained in the former. Therefore, the reason must be related to an efficient interadsorbate energy exchange occurring in the CO crowded coverages but not in the case of the CO 0.25~ML that characterized the low coverage.

\begin{figure}
\includegraphics*[angle=0,width=0.9\columnwidth]{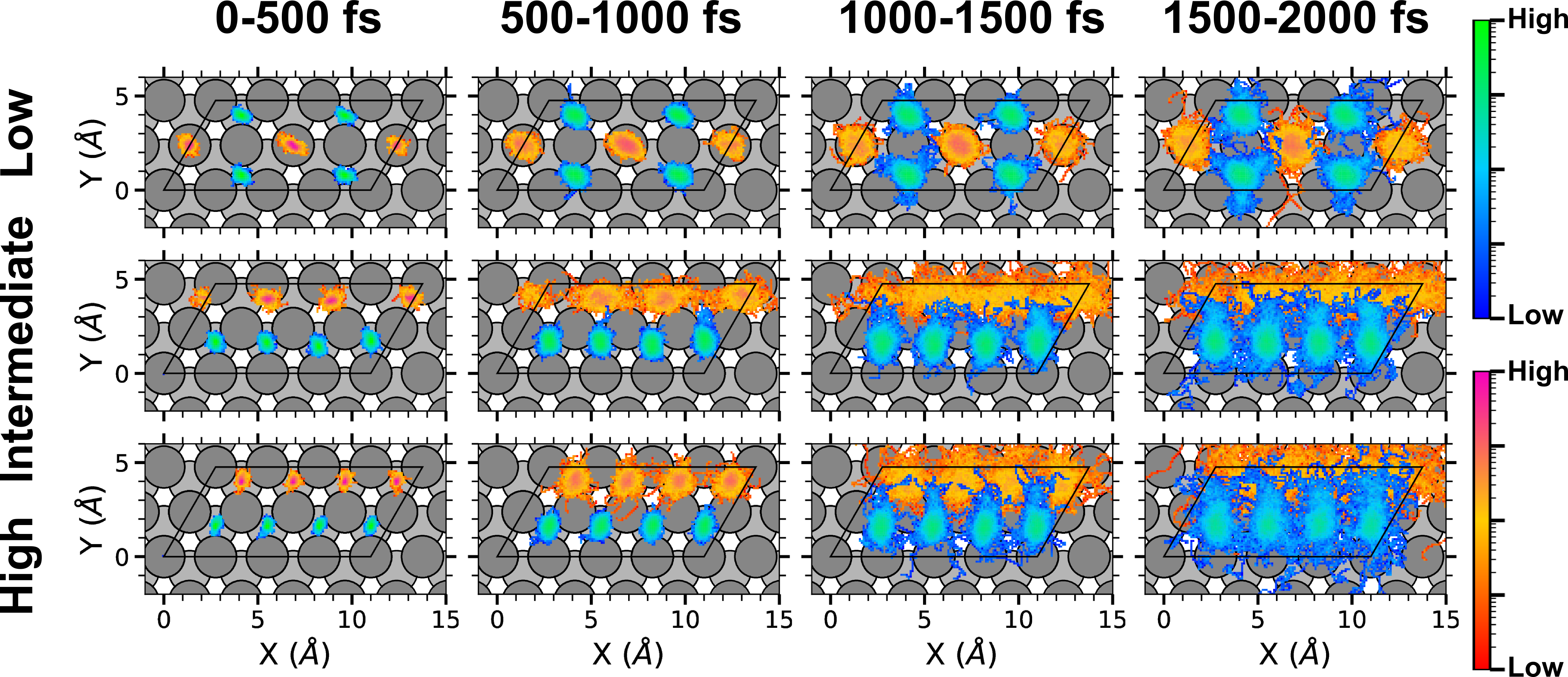}%
\caption{Two-dimensional histograms showing the time evolution of the CO center of mass position (orange) and O atom position (blue) on the surface for the low (top panels), intermediate (middle panels), and high (bottom panels) coverages.  
}
\label{fig:xy_CO}
\end{figure}

The two-dimensional (2D) histograms of Figure~\ref{fig:xy_CO}, showing the CO center of mass positions over the surface plane during the initial 2~ps, support the above idea. All the adsorbates remain in their respective adsorption wells in the interval $t<$0.5~ps. Thus, the dynamics in this interval is ruled by the coupling to the hot electrons as above stated. The differences in the CO dynamics start to be apparent in the interval $0.5$~ps$<t<$1.0~ps. A large fraction of the CO abandons the adsorption well in the intermediate coverage and to a lesser extent in the high coverage, while in the low coverage all CO still stay in the well atop a Ru atom. During the next picosecond, the CO molecules start to progressively leave the well in the low coverage, but during the same interval all the CO are already freely moving on the surface in the intermediate and high coverages. Note that in contrast to the intermediate and high coverages, there is a lack of intersections between the spots occupied by the CO molecules in the low coverage that support that the interadsorbate energy exchange is relevant in the two former coverages but not in the latter. Finally, both the slightly stronger coupling to the hot electrons and slightly larger energy exchange in the high coverage can explain the slight differences  between the high and intermediate coverages in $\langle E_\textrm{kin} \rangle(t)$.

Figure~\ref{fig:xy_CO} also provides indirect information on the properties of the potential energy surface (PES) for each coverage. An almost radial symmetry of the low coverage PES respect to the CO center of mass can be inferred from the observed symmetric CO $(X_\textrm{cm},Y_\textrm{cm})$-displacements and it is in accordance with the six-fold symmetry of the CO atop adsorption site. The PES dependence on the $(x_,y)$ coordinates of both the fcc and hcp O adsorbates has the (local) three-fold symmetry associated to both hollow sites, the minimum energy barriers for diffusion being located at the bridge sites. At $t$=1.0-2.0~ps the fraction of adsorbates reaching bridge positions is larger for O$_\textrm{fcc}$ than for O$_\textrm{hcp}$, i.e., for the adsorbates with the smallest desorption energy (0.643~eV against 1.206~eV)\cite{tetenoire2021}. The p(1$\times$2) arrangement of the intermediate coverage has low symmetry and this is clearly reflected in the PES properties. The histograms of the CO center of mass positions at $t<$1~ps show that these adsorbates, which are initially at nonequivalent near-top sites, can easily move along the $x$-axis but they are initially confined along the $y$-axis in between the two nearest rows of Ru atoms. At $t>$1 ps, the molecules start to have the necessary energy to also overcome these energy barriers. Interestingly, the local three-fold symmetry that is preserved in the honeycomb arrangement of the O-0.5~ML is broken in the O-p(1$\times$2) arrangement due to a non-negligible interaction between nearest O and CO adsorbates. The elongated shape of the corresponding 2D histograms along the $y$ direction suggests that the energy barriers at bridge sites between CO and O are smaller than the barriers at the bridge sites in between two O atoms. A similar PES dependence on the O $(x,y)$ coordinates is observed in the high-coverage PES, which is characterized by the same O-p(1$\times$2) arrangement. Differences appear when comparing the PES dependence on the CO positions. At $t$=0.5-1.0~ps, the high-coverage PES exhibits a local three-fold symmetry that agrees with the initial hcp sites occupied by all the CO molecules. Comparison of the histograms at $t$=1.0-1.5~ps between the high and intermediate coverages suggests that diffusion along the $x$ direction must be characterized by larger energy barriers in the CO-saturated high coverage, possibly caused by the more repulsive dipole-dipole interaction. 

\begin{figure}
\includegraphics*[angle=0,width=0.5\columnwidth]{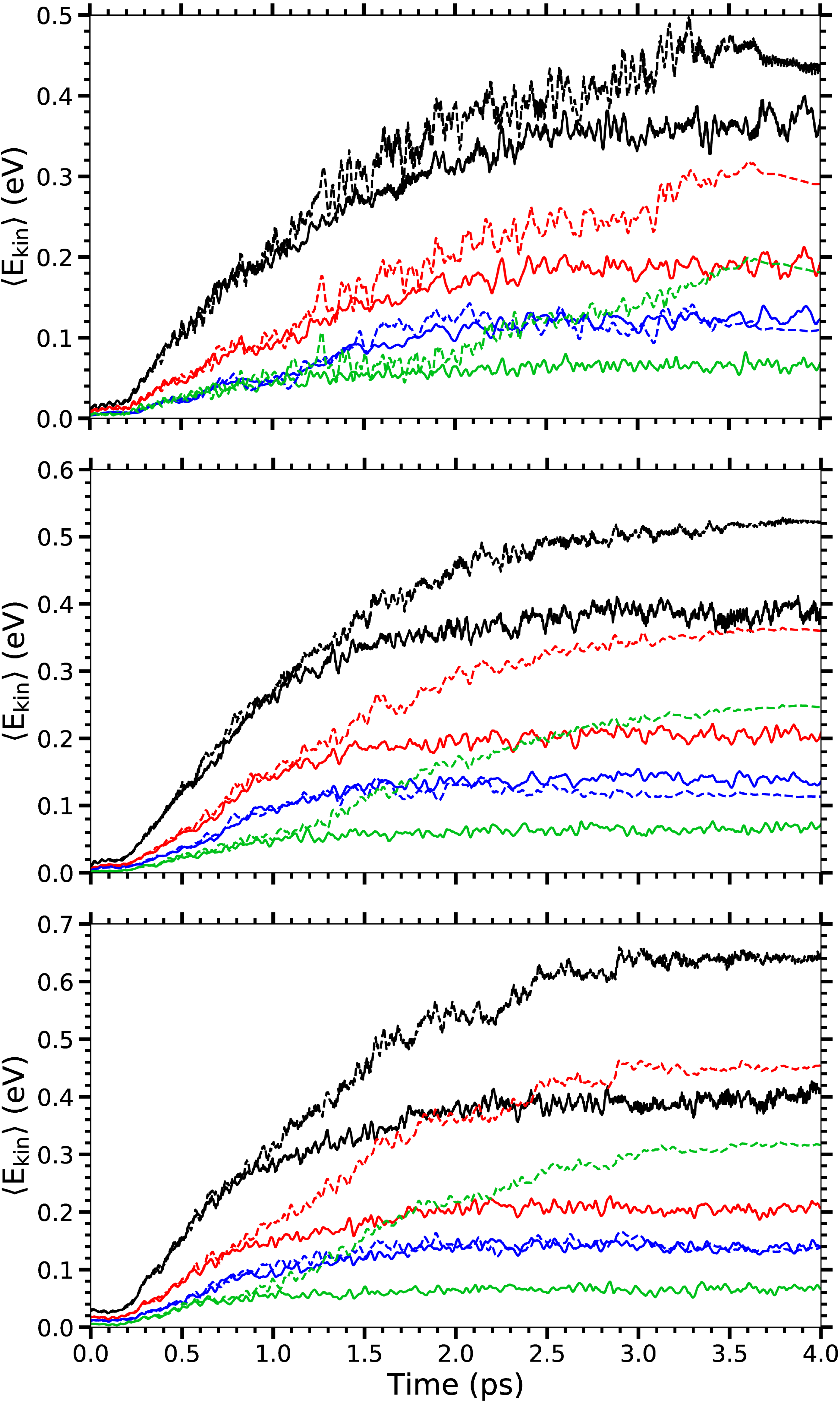}%
\caption{Mean kinetic energy of non-desorbing (solid lines) and desorbing (dashed lines) CO molecules against time for low (top panel), intermediate (middle panel), and high (bottom panel) coverages. Shown for each coverage are: the mean total kinetic energy $\langle E_\textrm{kin}\rangle$ (black), mean center-of-mass kinetic energy $\langle E_\textrm{cm}\rangle$ (red), and mean center-of-mass kinetic energy parallel (blue) and perpendicular (green) to the surface, $\langle E_\textrm{cm}^{\textrm{par}}\rangle$ and $\langle E_\textrm{cm}^{\textrm{z}}\rangle$, respectively.
}
\label{fig:ekin_des_ads}
\end{figure}

Finally, it remains to understand what properties determine the CO desorption dynamics. To this aim, we have compared different properties of the desorbing and non-desorbing CO molecules. The 2D histograms showing the time evolution of the $(X_\textrm{cm},Y_\textrm{cm})$ center of mass position for the desorbing (not shown) and non-desorbing (Figure~\ref{fig:xy_CO}) CO molecules are similar. The latter suggests that the configurational space leading to CO desorption is rather ample as it does not present special spatial features. A different initial coupling between desorbing and non-desorbing CO with the hot electrons can also be discarded because all the adsorption sites are respectively equivalent in the low (top sites) and high (hcp sites) coverages and, in the intermediate coverage, we verified that the desorption probability is unaffected by the slight differences between the three initial adsorption sites (CO1, CO2, and CO3 in Figure~\ref{fig:cell}), since the site-resolved desorption probabilities are basically the same for the three distinct CO. We do observe differences when comparing the mean kinetic energies $\langle E_\textrm{kin} \rangle (t)$. Figure~\ref{fig:ekin_des_ads} shows that at a certain instant (different for each coverage) the desorbing CO molecules (dashed lines) start to have more kinetic energy than the non-desorbing (solid lines) molecules (compare for each coverage the black dashed to the black solid lines). Note also that the extra energy gained by the desorbing CO mainly goes to the center-of-mass degrees of freedom. This can be inferred from the figure because the difference between the mean total kinetic energy of the desorbing and non-desorbing CO (i.e., difference between black dashed and black solid lines) is quite similar to the difference between the mean center-of-mass kinetic energy of the desorbing and non-desorbing CO (i.e., difference between red dashed and red solid lines). By decomposing the mean center-of-mass kinetic energy into parallel ($\langle E_\textrm{cm}^\textrm{par}\rangle (t)$, in blue) and perpendicular to the surface ($\langle E_\textrm{cm}^z\rangle (t)$, in green) contributions, it is clear that the desorbing molecules are characterized by having the largest $\langle E_\textrm{cm}^z\rangle (t)$ (green dashed lines), as one would naively expect. Interestingly, the parallel contribution of the desorbing and non-desorbing molecules is basically identical for each coverage. Altogether, we conclude that the desorbing molecules are characterized by a larger net energy gain that basically goes into the center-of-mass perpendicular motion.
Note in passing that at the end of the simulations, the relationships between the kinetic energies for the non-desorbing CO verify roughly equipartition among the different degrees of freedom, i.e., $\langle E_\textrm{kin} \rangle$:$\langle E_\textrm{cm}^\textrm{par}\rangle$:$\langle E_\textrm{cm}^z\rangle\equiv$ 6:2:1, that suggest a system arriving to a quasithermalized state at the end of the simulations. %

\subsection{CO Oxidation Dynamics}

The number of oxidation events in our simulations is very scarce which, in principle, represents a question mark about the statistical validity of the results. However, taking this limitation in mind, the following observations on the differences found on the oxidation dynamics at the different coverages look very general and robust. The instants at which CO$_2$ desorbs from each covered surface are compared in Figure~\ref{fig:2tm} (top panel). The two oxidation events obtained in the low coverage occur at well separated instants, $t\sim$2 and 4~ps, respectively. In contrast, in the high and intermediate coverages, all the CO$_2$ desorption events occur within  a shorter time interval of about 1~ps. The important difference between these two coverages is that the first event takes place at $t\sim$1.4~ps in the former and at $t\sim$2.6~ps in the latter. Considering that the high coverage has the largest activation energy for CO oxidation, 
not only is it surprising to find that the oxidation probability is the highest for this coverage (see Table~\ref{tab:table1}) but also that the process initiates more rapidly on this surface. There are various reasons that can explain this unexpected result. As above discussed, the adsorbates gain more energy and at faster rate in the high coverage because the coupling to the excited electrons is initially the strongest and, importantly,  because the interadsorbates energy exchange is more efficient in this crowded surface. Furthermore, we cannot discard that the energy barriers on the highly excited surface are different from those calculated on the relaxed surface. In this respect, the analysis of the CO oxidation trajectories below allows us to determine up to what extent the followed oxidation paths coincide with the minimum reaction paths we identified for each coverage in our previous static calculations~\cite{tetenoire2021}. 

As already discussed in ref~\citenum{tetenoire2022}, the two CO oxidation trajectories in the low coverage follow the corresponding DFT+vdW-DF minimum energy path (MEP)~\cite{tetenoire2021}, in which the less bound O$_\textrm{fcc}$ surmounts its adsorption well, crosses the bridge site between two Ru atoms, and recombines with the nearby CO that tilts to form the chemisorbed bent CO$_2$ (bCO$_2$). However, compared to the low coverage, not only the adsorbates become more rapidly very mobile in the intermediate and high coverages (see Figure~\ref{fig:xy_CO}), but also the Ru atoms in the two topmost layers (not shown). As a result, none of the oxidation trajectories seem to follow the MEP calculated for their corresponding equilibrium structures. The latter explains why the CO$_2$ desorption probabilities are not necessarily in accordance to the activation energies calculated for the relaxed structures. Furthermore, the oxidation process in the intermediate coverage and, particularly, in the high coverage is occasionally preceded by other desorption events that contribute to reduce the nominal CO coverage and, concomitantly, to change the activation energies, remarking altogether the importance of the reaction dynamics. 

\begin{figure}
\includegraphics*[angle=0,width=0.9\columnwidth]{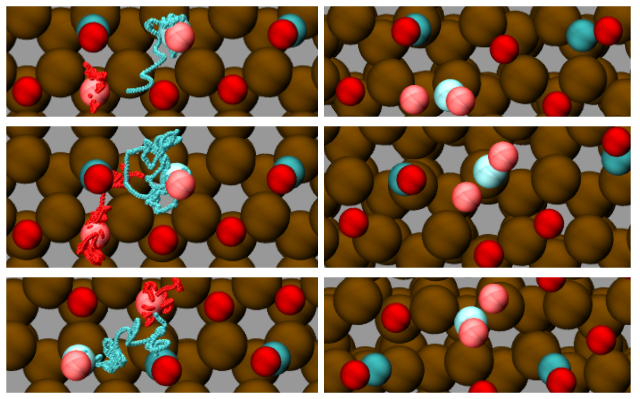}%
\caption{Examples of three (out of the four) CO oxidation trajectories obtained in the intermediate coverage from the ($T_\textrm{e}, T_\textrm{l}$)-AIMDEF simulations (brown, red, and blue spheres correspond to Ru, O, and C atoms, respectively). For each trajectory the initial position of the adsorbates and Ru atoms is depicted in the left panel together with the evolution of the reactive O (red thick lines) and CO (blue thick lines), while the right panel shows the atoms positions at the instant at which the bCO$_2$ formation starts. For clarity, in all the panels the reactive O and CO are depicted by light red and light blue spheres, respectively. Images prepared with the VMD software~\cite{vmd}.
\label{fig:oxidation-intermediate}}
\end{figure}
\begin{figure}
\includegraphics*[angle=0,width=0.9\columnwidth]{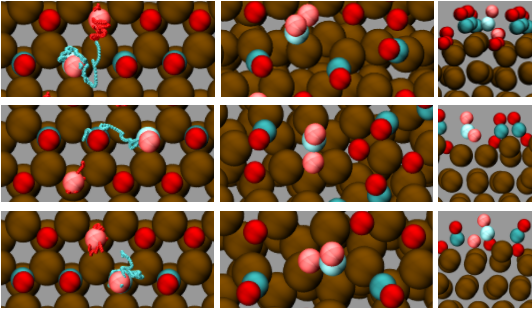}%
\caption{Examples showing the variety of CO oxidation trajectories that are obtained in the high coverage from the ($T_\textrm{e}, T_\textrm{l}$)-AIMDEF simulations. For each trajectory the initial position of the adsorbates and Ru atoms is depicted in the left panel together with the evolution of the reactive O (red thick lines) and CO (blue thick lines), while the middle and right panels show, respectively, top and side views of the atoms positions at the instant at which the bCO$_2$ formation starts. Color code as in Figure~\ref{fig:oxidation-intermediate}. Images prepared with the VMD software~\cite{vmd}.
\label{fig:oxidation-high}}
\end{figure}

The trajectories depicted in Figures~\ref{fig:oxidation-intermediate} and \ref{fig:oxidation-high} for the intermediate and high coverages, respectively, are good examples of the variety of the reaction paths that, at variance with the calculated MEP, do not necessarily involve recombination of nearest adsorbates, since in most cases the reactive species diffuse on the surface prior to recombining into CO$_2$. Figure~\ref{fig:oxidation-intermediate} shows the three trajectories in the intermediate coverage not involving recombination between nearest neighbors. Note that in two of them (top and bottom panels), diffusion does not involve the reacting O adsorbate, as obtained in the MEP calculation for this coverage, but the reacting CO molecule. In the middle panel, O diffuses, but at variance with the MEP, it recombines with a second nearest CO neighbor.

In the high coverage, all the recombinations except one (the one shown in the middle row) take place between nearest neighbors. Remarkably, a common event occurring in the high coverage that is favoring the formation of the chemisorbed bCO$_2$, and subsequent CO$_2$ desorption, is the pronounced outwards displacement of the Ru atom that binds the reactive CO. This initial distortion may eventually end in the formation of a quasi-stepped Ru topmost layer with the O atoms and the CO molecules adsorbed in the lowest and the highest Ru rows, respectively. This arrangement in which CO adsorbs at a higher height and on a low coordinated site facilitates the bCO$_2$ formation. It resembles the transition state geometries found in diverse stepped surfaces, although in those cases it is the reactive O and not the CO molecule that adsorbs at the low coordinated site\cite{kim07,zeljko10,zhou19,zhao22}. Figure~\ref{fig:oxidation-high} shows examples of these two types of surface distortions. The trajectory depicted in the top panels corresponds to the mentioned stepped-like distortion that we observe in one of the oxidation events. The side view plotted in the top-right panel shows clearly that the whole Ru row supporting the CO adsorbates is shifted upwards, leaving the O adatoms conveniently located at lower heights to facilitate formation of the O-CO bond. The trajectory in the bottom panels corresponds to the case in which only the Ru bound to the reactive CO is upwards displaced, as can be observed in the side view depicted in the bottom-right panel. The trajectory in the middle panels is an example in which the distorted Ru atom is not directly involved in the recombinative CO$_2$ desorption. Altogether the above analysis emphasizes that understanding the adsorbates reactivity under these extremely non-equilibrium conditions requires an adequate treatment of the adsorbates dynamics and also of the surface atoms dynamics. Not only the important coupling to the laser-excited electrons, but also other factors such as the interadsorbate energy exchange, the eventual high lattice temperatures, and additional surface distortions created by the highly excited adsorbates can contribute to the final photo-induced reactivity on surfaces.

\section{Conclusions \label{sec:conclusions}}
In summary, we have studied the photo-induced desorption and oxidation of CO in different (O, CO)-covered Ru(0001) surfaces, using ab initio molecular dynamics simulations that describe the laser-excited system in terms of non-equilibrated time-dependent electronic and phononic temperatures (($T_\textrm{e}, T_\textrm{l}$)-AIMDEF). The simulated coverages are compatible with the preparation of the (O,CO) overlayer in existing femtosecond laser experiments aimed to understand the competition between CO desorption and oxidation on the Ru(0001) surface~\cite{bonn99,obergJCP2015}. Thus, coadsorbed with the oxygen saturation coverage of 0.5~ML we have considered the following coverages: (i) the low coverage of 0.5ML O+0.25ML CO, (ii) the intermediate coverage of 0.5ML O+0.375ML CO, and (iii) the high coverage of 0.5ML O+0.5ML CO. In the three cases, our ($T_\textrm{e}, T_\textrm{l}$)-AIMDEF simulations yield CO desorption to oxidation branching ratios larger than one order of magnitude, which is consistent with the experimental findings. Interestingly, these large branching ratios cannot be explained in terms of the difference between the activation energies for CO desorption and CO oxidation of a few hundreds of meV ($E_\textrm{TS}$(CO)-$E_\textrm{TS}$(CO$_2$)$\simeq-0.4$, 0.2, and 0.1 eV for low, intermediate, and high coverages, respectively). The analysis of the photo-induced CO desorption and oxidation dynamics allows us to understand the coverage dependence of both processes. 

We observe that the configurational space leading to CO desorption is quite ample. As a consequence, the CO desorption is a rather simple and direct process only limited by the energy the CO molecules need to gain in order to overcome the energy barrier to desorption. This is reflected in the dependence of the CO desorption probability on coverage. This probability is the largest for the intermediate coverage, which has the lowest energy barrier to desorption (around 0.6~eV), and it is the smallest for the low coverage, which presents the highest energy barrier to desorption (1.57~eV). By comparing the time evolution of the desorption probabilities, we identify other factors that also contribute to the dependence of the desorption dynamics on coverage, the coupling to the electronic system and the interadsorbate energy exchange. The former, being quite similar for the three coverages, is only responsible of the slightly larger energy that the CO molecules gain initially in the high coverage. The efficient interadsorbate energy exchange that is taking place in the high and intermediate coverages explains the faster energy uptake and faster desorption dynamics observed in these coverages as compared to the low coverage. In the three cases, the desorbing CO are characterized by gaining more energy than the non-desorbing CO. The extra energy is mainly deposited in the translational degree of freedom normal to the surface.

In contrasts to CO desorption, the CO oxidation probabilities do not correlate with the values of the activation energies for CO oxidation that were calculated in the relaxed surface. 
For instance, the CO oxidation probability is the highest for the high coverage, i.e., the coverage with the largest activation energy. Our dynamics simulations show that CO oxidation in these nonequilibrium conditions presents a much more complex scenario than CO desorption. Except for the low coverage case, the oxidation events taking place in our dynamics do not follow the minimum energy paths calculated under equilibrium conditions. Moreover, oxidation is sometimes preceded in the intermediate and high coverages by CO desorption events that modify the local environment. The dynamics simulations also show that the excited adsorbates, that become particularly mobile in the intermediate and high coverages, can cause profound surface distortions. In particular, a common event occurring in the high coverage that is favoring the formation of the chemisorbed bent CO$_2$, and subsequent
CO$_2$ desorption, is the pronounce outwards displacement of the Ru atom that binds the
reactive CO. As a consequence of all it, the relevant energy barriers for the oxidation process in the highly excited surface are different from those calculated on the relaxed surface. All in all, these results show the importance of accounting fot the adsorbate and surface dynamics under extremely non-equilibrium conditions in order to understand the laser induced photochemistry at metal surfaces. 

\begin{acknowledgement}
The authors acknowledge financial support by the Gobierno Vasco-UPV/EHU
Project No.  IT1569-22 and by the Spanish MCIN/AEI/10.13039/501100011033 [Grant No. PID2019-107396GB-I00]. This research was conducted in the scope of the Transnational Common Laboratory (LTC) “QuantumChemPhys – Theoretical Chemistry and Physics at the Quantum Scale”. Computational resources were provided by the DIPC computing center.
\end{acknowledgement}




\bibliography{refs-acs}

\end{document}